\documentclass[12pt,oneside]{article}
\usepackage[T1]{fontenc}
\usepackage{enumitem}
\usepackage[utf8]{inputenc}
\usepackage{babel}
\usepackage[tt=lining]{cfr-lm}
\usepackage[hyphens]{url} 
\usepackage{amsmath,amsthm}	% Advanced maths commands
\usepackage{amssymb}	% Extra maths symbols
\usepackage{mathtools} % Bonus

\begin{document}

\begin{center}
\vskip 1cm{\LARGE\bf
Evolution of Curvature in Riemannian \vskip 0.2cm Geometry
}
\vskip 1cm
\large
Abhishek Das \\
B. M. Birla Science Centre \\
Adarsh Nagar, Hyderabad \\
India, 500063 \\
{\tt abhik.das31@gmail.com} \\
\vskip 1cm
\end{center}
\begin{abstract}
In this paper we shall follow two different routes that embody the existence of a point of \emph{exclusivity} - the opposite of a \emph{singularity}. First, we show that this follows from Raychaudhuri's equation. Then, we substantiate that the evolution of the Riemann-Christoffel tensor can be expressed entirely by an arbitrary timelike vector field and that the curvature tensor returns to its initial value with respect to change in a particular index. It has been shown that geodesics can diverge just as they can converge, resulting in a point of \emph{exclusivity} and point of \emph{singularity}.\\
\end{abstract}

\section{Introduction}
General theory of relativity, the geometric theory of gravitation is based on the plinth of Riemannian geometry that studies differentiable manifolds. There is a plethora of literature on these two intricately related topics. As is known, there are a multitude of implications resulting from Riemannian geometry when applied to the general relativity - the Einstein field equations \cite{Einstein}, solutions to such equations in the form of Schwarzchild metric \cite{Karl}, Friedmann-Lemaitre-Robertson-Walker metric \cite{Friedmann1,Friedmann2,Lemaitre1,Lemaitre2,Robertson1,Robertson2,Robertson3,Walker} and others \cite{Kerr,Ezra}, singularities \cite{Raychaudhuri} and black holes \cite{Penrose1,Penrose2}.\\
However, in essence, the current paper is not innately related to the general theory of relativity. It is essentially devoted to the geometric aspect of space-time, somewhat in contradistinction to a recent paper that was based on a Hamiltonian formulation \cite{Das}. The inception begins with the feasibility of existence of a {\it point of exclusivity} (the opposite of a {\it singularity}), that is corroborated considering the Raychaudhuri equation \cite{Raychaudhuri}. And then, we consider parallel transported vectors and thereby manipulating of their connection with the geometry of space-time we are led to an evolution equation for the Riemann-Christoffel tensor. As a consequence, several interesting implications are drawn out. Most importantly, it has been argued that with necessary conditions fulfilled one can indeed discern the existence of a {\it point of exclusivity}. \\

\section{The positive and negative values of the expansion scalar}
In his seminal paper of 1955 \cite{Raychaudhuri}, Raychaudhuri had obtained an equation that led to the focusing theorem which in turn substantiated the existence of singularities and black holes, eventually. We would find in the present section that the result derived by Raychaudhuri plays a significant role in some novel aspects. Let us commence with the expansion scalar ($\theta$) of the Raychaudhuri equation. We know that\\
\[\theta = \frac{\partial}{\partial t}(\ln G)\]\\
where, $G = \sqrt{-g}$. Also, we know that\\
\[A_{;i}^{i} = \frac{1}{\sqrt{-g}}[A^{i}\frac{\partial}{\partial x^{i}}\sqrt{-g} + \sqrt{-g}
\frac{\partial}{\partial x^{i}}A^{i}]\]\\
So, considering the coordinate $t$ ($i = 0$) we may write\\
\[A_{;0}^{0} = [A^{0}\theta + A^{0}_{,0}]\]\\
Now, since the Kronecker delta function is independent of both the normal and covariant derivatives the last equation can be equivalently expressed as\\
\begin{equation}
A_{k;0} = A_{k}\theta + A_{k,0} \label{31}
\end{equation}
Again, we also know that\\
\[A_{k;i} = A_{k,i} - \Gamma_{ki}^{l}A_{l}\]\\
Therefore, using equation (\ref{31}) we obtain\\
\begin{equation}
A_{k}\theta + \Gamma_{k0}^{l}A_{l} = 0 \label{32}
\end{equation}
or,\\
\[A_{k}\theta + \Gamma_{k0}^{l}\delta_{l}^{k}A_{k} = 0\]\\
Since, the vector fields $A_{k}$'s are arbitrary, we have the following equation for the expansion scalar\\
\begin{equation}
\theta + \Gamma_{k0}^{l}\delta_{l}^{k} = 0 \label{33}
\end{equation}
One can make two immediate observations from equation (\ref{33}). If $k \neq l$, we have\\
\begin{equation}
\theta = 0 \label{34}
\end{equation}
And, if $k = l$, we have\\
\begin{equation}
\theta + \Gamma_{l0}^{l} = 0 \label{35}
\end{equation}
Again, on account of the relation\\
\[\Gamma_{l0}^{l} = \frac{1}{\sqrt{-g}}\frac{\partial}{\partial x^{0}}\sqrt{-g} = \frac{\partial}{\partial x^{0}}(\ln \sqrt{-g})\]\\
we derive from equation (\ref{35})\\
\begin{equation}
\theta + \theta = 0
\end{equation}
which is, in essence, similar to equations (\ref{26}). Either the expansion scalar is zero, or\\
\begin{equation}
|\theta| = \pm \theta \label{37}
\end{equation}
which seems erroneous from a mathematical perspective. But, suppose $\theta$ can actually have the same value with opposite signs; then the last two equations imply some physical meaning. This shall be further corroborated by the following methodology. Equation (\ref{32}) can also be written as (with, $A_{0}^{k} = \Gamma_{k0}^{l}A_{l}$)\\
\[A_{k}\theta + A_{0}^{k} = 0\]\\
Differentiating with respect to $x^{0}$  and using the above equation again we have\\
\begin{equation}
A_{k}\dot{\theta} - \frac{A_{k,0}A^{k}A^{k}_{0}}{A^{2}} + A^{k}_{0,0} = 0 \label{38}
\end{equation}
where, $A^{2} = A^{k}A_{k}$. Again, Raychaudhuri's equation is as follows:\\
\begin{equation}
\dot{\theta} = \frac{\theta^{2}}{3} - 2\sigma^{2} + 2\omega^{2} - R_{s} \label{39}
\end{equation}
where, the symbols have their usual meanings. Thus, from equations (\ref{38}) and (\ref{39}) we have\\
\[\frac{A^{2}\theta^{2}}{3} + A^{2}(2\omega^{2} - \sigma^{2} + R_{s}) + A_{k,0}A^{k}\theta + A^{k}A^{k}_{0,0} = 0\]\\
Writing $\epsilon = 2\omega^{2} - 2\sigma^{2} - R_{s}$, and solving the quadratic equation we finally derive\\
\begin{equation}
\theta = \frac{3}{2}\left[-\frac{A_{k,0}A^{k}}{A^{2}} \pm \sqrt{\frac{(A_{k,0}A^{k})(A_{k,0}A^{k})}{A^{4}} - \frac{4}{3}\left\{\frac{A_{k,0}A^{k}}{A^{2}} + \epsilon\right\}}\right]
\end{equation}
Here, we can derive some interesting conclusions. When the discriminant of the above solution is zero, we would have another quadratic equation of the form:\\
\[p^{2} - \frac{4}{3}p - \frac{4}{3}\epsilon = 0\]\\
where, $p = \frac{A_{k,0}A^{k}}{A^{2}}$. The solution is of the form\\
\[p = \frac{2}{3} \pm \frac{2}{3}\sqrt{1 - 3\epsilon}\]\\
which implies that $\epsilon \geq \frac{1}{3}$. On the other hand, if we have\\
\[\frac{(A_{k,0}A^{k})(A_{k,0}A^{k})}{A^{4}} = \frac{4}{3}\left\{\frac{A_{k,0}A^{k}}{A^{2}} + \epsilon\right\}\]\\
then, in such a scenario\\
\[\theta = -\frac{3}{2}\frac{A_{k,0}A^{k}}{A^{2}}\]\\
This may be considered the initial value of $\theta$ during the birth of the universe when all geodesics were focused at a particular point and then after the {\it big bang} as the universe began to evolve the value of $\theta$ evolved - accumulating the discriminant term. Again, if the Raychaudhri scalar is such that\\
\[R_{s} > \frac{A_{k,0}A^{k}}{A^{2}} + 2\omega^{2} - 2\sigma^{2}\]\\
then\\
\[\sqrt{\frac{(A_{k,0}A^{k})(A_{k,0}A^{k})}{A^{4}} - \frac{4}{3}\left\{\frac{A_{k,0}A^{k}}{A^{2}} + \epsilon\right\}} > \frac{A_{k,0}A^{k}}{A^{2}}\]\\
So, $\theta$ can have positive values as well. Therefore, writing 
\[\varepsilon = -\frac{A_{k,0}A^{k}}{A^{2}} \pm \sqrt{\frac{(A_{k,0}A^{k})(A_{k,0}A^{k})}{A^{4}} - \frac{4}{3}\left\{\frac{A_{k,0}A^{k}}{A^{2}} + \epsilon\right\}}\]\\
we would have\\
\begin{equation}
\theta = \pm \frac{3}{2}\varepsilon \label{40}
\end{equation}
This is what insinuates from equation (\ref{37}) too. Hence, essentially, $\dot{\theta}$ can diverge to both positive and negative infinity. The physical significance of this conclusion is novel and significant - geodesics can diverge just as they can converge. As a consequence, there will exist a point of exclusivity akin to a point of singularity. This point of exclusivity can be looked upon as the origin of all matter and energy and the formation of spacetime as we know it. We understand immdiately how important the Raychaudhuri scalar and expansion scalar are.\\
However, it should be borne in the mind that the notion and methodology innovated here is in stark contrast with the theory of {\it Big Rip} \cite{Caldwell}, particularly for the fact that no speculative, mysterious energy has been talked about - here it is essentially the geometry trying to explain the evolution of the universe.\\

\section{The evolution equation}
Now, we shall substantiate the novel result of the preceding section taking a different route. At first, let us consider a covariant vector $B_{i}$ whose transport between two different points of a Riemannian manifold is independent of the path and that is not covariantly constant; then it is known that the derivative of this vector field would be given as \cite{Jayant}\\
\begin{equation}
\frac{\partial B_{i}}{\partial x^{k}} = \Gamma_{ik}^{l}B_{l} \label{1}
\end{equation}
We assume that $B_{i}$ is thrice differentiable. Thus we have the following from relation (\ref{1})
\begin{equation}
\frac{\partial^{2} B_{i}}{\partial x^{n}\partial x^{k}} = \frac{\partial\Gamma_{ik}^{l}}{\partial x^{n}}B_{l} + \Gamma_{ik}^{l}\frac{\partial B_{l}}{\partial x^{n}} \label{2}
\end{equation}
and\\
\begin{equation}
\frac{\partial^{3} B_{i}}{\partial x^{m}\partial x^{n}\partial x^{k}} = B_{i,mnk} =  \frac{\partial^{2}\Gamma_{ik}^{l}}{\partial x^{m}\partial x^{n}}B_{l} + \Gamma_{ik}^{l}\frac{\partial^{2} B_{l}}{\partial x^{m}\partial x^{n}} + \frac{\partial\Gamma_{ik}^{l}}{\partial x^{n}}\frac{\partial B_{l}}{\partial x^{m}} + \frac{\partial\Gamma_{ik}^{l}}{\partial x^{m}}\frac{\partial B_{l}}{\partial x^{n}} \label{3}
\end{equation}
Now, computing $B_{i,mkn}$, subtracting the resultant equation from (\ref{3}) and then changing the indices as $l \rightarrow j$, we have the equation\\
\begin{equation}
B_{i,mnk} - B_{i,mkn} = \Gamma_{ik,mn}^{j}B_{j} - \Gamma_{in,mk}^{j}B_{j} + \Gamma_{ik}^{j}B_{j,mn} - \Gamma_{in}^{j}B_{j,mk} + \theta \label{4}
\end{equation}
where, $\theta = \Gamma_{ik,n}^{j}B_{j,m} + \Gamma_{ik,m}^{j}B_{j,n} - \Gamma_{in,k}^{j}B_{j,m} - \Gamma_{in,m}^{j}B_{j,k}$ and the 'comma' implies normal derivative. Now, we know the following relations regarding the Riemann-Christoffel (RC) tensor\\
\[B_{i;kn} - B_{i,nk} = -R_{k~ni}^{m}B_{m}\]\\
and\\
\[B_{i;nk} - B_{i,kn} = -R_{n~ki}^{m}B_{m}\]\\
where, the 'semi-colon' implies covariant derivative. Therefore, we can write the following from equation (\ref{4})\\
\[\partial_{m}[B_{i,nk} - B_{i,kn}] = \partial_{m}[2R_{i~kn}^{j}B_{j} + \eta B_{i}] = \Gamma_{ik,mn}^{j}B_{j} - \Gamma_{in,mk}^{j}B_{j} \Gamma_{ik}^{j}B_{j,mn} - \Gamma_{in}^{j}B_{j,mk} + \theta\]\\
where, $\eta B_{i} = B_{i;kn} - B_{i;nk}$ ($\eta$ being a covariant derivative operator). This yields the premature evolution equation for the RC tensor or the curvature tensor as\\
\begin{equation}
2\partial_{m}[R_{i~kn}^{j}]B_{j} + 2R_{i~kn}^{j}B_{j,m} + [\Gamma_{in,mk}^{j} - \Gamma_{ik,mn}^{j}]B_{j} + \Gamma_{in}^{j}B_{j,mk} -  \Gamma_{ik}^{j}B_{j,mn} + \partial_{m}(\eta B_{i}) - \theta = 0\label{5}
\end{equation}
It is worth noting that the vector fields $B_{i}$ can be related to all the necessary features of the manifold and do not depend explicitly on the Christoffel symbols, in this regard. Now, with the differential equation\\
\[\frac{\partial B_{i}}{\partial x^{n}} = \Gamma_{in}^{m}B_{m}\]
we shall have equation (\ref{2}) as\\
\[B_{i,kn} = \Gamma_{in,k}^{m}B_{m} + \Gamma_{in}^{m}\Gamma_{mk}^{p}B_{p}\]\\
Differentiating this again and rearranging we get\\
\[\Gamma_{in,jk}^{m}B_{m} = B_{i,jkn} - \Gamma_{in,k}^{m}\Gamma_{mj}^{r}B_{r} - \Gamma_{in}^{m}\Gamma_{mk,j}^{p}B_{p} - \Gamma_{in,j}^{m}\Gamma_{mk}^{p}B_{p} - \Gamma_{in}^{m}\Gamma_{mk}^{p}B_{p,j}\]\\
Now, interchanging the indices $m$ and $j$ ($m \leftrightarrow j$) we have\\
\begin{equation}
\Gamma_{in,mk}^{j}B_{j} = B_{i,mkn} - \Gamma_{in,k}^{j}\Gamma_{jm}^{r}B_{r} - \Gamma_{in}^{j}\Gamma_{jk,m}^{p}B_{p} - \Gamma_{in,m}^{j}\Gamma_{jk}^{p}B_{p} - \Gamma_{in}^{j}\Gamma_{jk}^{p}B_{p,m} \label{6}
\end{equation}
Similarly, we would have\\
\begin{equation}
\Gamma_{ik,mn}^{j}B_{j} = B_{i,mnk} - \Gamma_{ik,n}^{j}\Gamma_{jm}^{r}B_{r} - \Gamma_{in}^{j}\Gamma_{jn,m}^{p}B_{p} - \Gamma_{in,m}^{j}\Gamma_{jn}^{p}B_{p} - \Gamma_{ik}^{j}\Gamma_{jn}^{p}B_{p,m} \label{7}
\end{equation}
Again, since $\Gamma_{in,k}^{j}B_{j} = B_{i,kn} - \Gamma_{in}^{j}\Gamma_{jk}^{p}B_{p}$, the equations (\ref{6}) and (\ref{7}) can be rewritten respectively as\\
\begin{equation}
\begin{aligned}
\Gamma_{in,mk}^{j}B_{j} &= B_{i,mkn} - \Gamma_{in}^{j}\Gamma_{jk}^{p}\Gamma_{pm}^{q}B_{q} - \delta_{r}^{j}\Gamma_{jm}^{r}\{B_{i,kn} - \Gamma_{in}^{j}\Gamma_{jk}^{p}B_{p}\}\\
&\quad - \Gamma_{in}^{j}\{B_{j,mk} - \Gamma_{jk}^{p}\Gamma_{pm}^{q}B_{q}\} - \delta_{p}^{j}\Gamma_{jk}^{p}\{B_{i,mn} - \Gamma_{in}^{j}\Gamma_{jm}^{p}B_{p}\} \label{8}
\end{aligned}
\end{equation}
and\\
\begin{equation}
\begin{aligned}
\Gamma_{ik,mn}^{j}B_{j} &= B_{i,mnk} - \Gamma_{ik}^{j}\Gamma_{jn}^{p}\Gamma_{pm}^{q}B_{q} - \delta_{r}^{j}\Gamma_{jm}^{r}\{B_{i,nk} - \Gamma_{ik}^{j}\Gamma_{jn}^{p}B_{p}\}\\
&\quad - \Gamma_{ik}^{j}\{B_{j,mn} - \Gamma_{jn}^{p}\Gamma_{pm}^{q}B_{q}\} - \delta_{p}^{j}\Gamma_{jn}^{p}\{B_{i,mk} - \Gamma_{ik}^{j}\Gamma_{jm}^{p}B_{p}\} \label{9}
\end{aligned}
\end{equation}
Now, subtracting equation (\ref{9}) from equation (\ref{8}) and rearranging, we have\\
\begin{equation}
\begin{aligned}
B_{j}[\Gamma_{in,mk}^{j} - \Gamma_{ik,mn}^{j}] &= (B_{i,mkn} - B_{i,mnk}) + \Gamma_{jm}^{j}(B_{i,nk} - B_{i, kn}) + \Gamma_{ik}^{j}B_{j,mn} - \Gamma_{in}^{j}B_{j,mk}\\
&\quad + \delta_{p}^{j}(\Gamma_{jn}^{p}B_{i,mk} - \Gamma_{jk}^{p}B_{i,mn}) + 2\delta_{p}^{j}B_{j,m}(\Gamma_{in}^{j}\Gamma_{jk}^{p} - \Gamma_{ik}^{j}\Gamma_{jn}^{p}) \label{10}
\end{aligned}
\end{equation}
Let us consider the second term with parenthesis on the right hand side of the last equation. Multiplying by $B^{2}B_{j} = B^{j}B_{j}B_{j} = B^{j}B^{2}_{j}$ we shall have\\
\[B^{2}B_{j}\Gamma_{jm}^{j}(B_{i,nk} - B_{i, kn}) = B^{j}B_{j}B_{j}\Gamma_{jm}^{j}(B_{i,nk} - B_{i, kn}) = B^{2}B_{j,m}(B_{i,nk} - B_{i, kn})\]\\
Clearly, there is a breakdown of index notation, pertinent to the index 'j'. We shall elaborate this scenario now and we shall find the above equation to be useful subsequently. We make an ansatz that in this special scenario, the first term in the parenthesis in equation (\ref{10}), namely $(B_{i,mkn} - B_{i,mnk})$, becomes explicitly independent of the index $j$ present in the left hand side. The rationale can be attributed to when some structure preserving endomorphism that preserves the geometry of the manifold breaks down and thereby the RC tensor accrues a new upper index and the index notation breaks down.\\
The feasibility of the rationale introduced above will be evident later while elucidating equation (\ref{19}). So, essentially, the RC tensor arising from the breakdown will be independent of the index $j$. This causes the breakdown of the 'index notation' mentioned earlier. Therefore, considering this ansatz we may write\\
\[B_{i,mkn} - B_{i,mnk} = -\partial_{m}[2R_{i~kn}^{p}B_{p} + \eta B_{i}] = -2\partial_{m}[R_{i~kn}^{p}]B_{p} - 2[R_{i~kn}^{p}]B_{p,m} - \partial_{m}(\eta B_{i})\]
where we have considered a new index - $p$, such that $j \neq p$. Thus, using (\ref{10}) and multiplying both sides of equation (\ref{5}) by $B^{2} = B^{j}B_{j} = \delta^{jj}B_{j}B_{j}$ we derive\\
\begin{equation}
\begin{aligned}
&\quad 2B^{j}B^{2}_{j}[\partial_{m}\{R_{i~kn}^{j}\}B_{j} - \partial_{m}\{R_{i~kn}^{p}\}B_{p}] + 2B^{2}B_{j}[\{R_{i~kn}^{j}\}B_{j,m} - \{R_{i~kn}^{p}\}B_{p,m}]\\
&\quad + B^{2}B_{j,m}(B_{i,nk} - B_{i, kn})
+ \delta_{p}^{j}B^{j}B^{2}_{j}(\Gamma_{jn}^{p}B_{i,mk} - \Gamma_{jk}^{p}B_{i,mn})\\
&\quad + 2\delta_{p}^{j}B^{j}B^{2}_{j}B_{j,m}(\Gamma_{in}^{j}\Gamma_{jk}^{p} - \Gamma_{ik}^{j}\Gamma_{jn}^{p}) - B^{j}B^{2}_{j}\theta = 0 \label{11}
\end{aligned}
\end{equation}
\\
\\
Now, we shall use the relation $\Gamma_{in,k}^{j}B_{j} = B_{i,kn} - \Gamma_{in}^{j}\Gamma_{jk}^{p}B_{p}$, and compute $\theta$ as follows:\\
\begin{equation}
\begin{aligned}
B_{j}^{2}\theta &= B_{j}B_{j,m}(B_{i,nk} - \Gamma_{ik}^{j}\Gamma_{jn}^{p}B_{p}) + B_{j}B_{j,n}(B_{i,km} - \Gamma_{im}^{j}\Gamma_{jk}^{p}B_{p})\\
&\quad - B_{j}B_{j,m}(B_{i,kn} - \Gamma_{in}^{j}\Gamma_{jk}^{p}B_{p}) - B_{j}B_{j,k}(B_{i,nm} - \Gamma_{im}^{j}\Gamma_{jn}^{p}B_{p})
\end{aligned}
\end{equation}
from which we obtain\\
\begin{equation}
\begin{aligned}
B_{j}^{2}\theta &= B_{j,m}(B_{j}B_{i,nk} - B_{i,k}B_{j,n}) + B_{j,n}(B_{j}B_{i,km} - B_{i,m}B_{j,k})\\
&\quad - B_{j,m}(B_{j}B_{i,kn} - B_{i,n}B_{j,k}) - B_{j,k}(B_{j}B_{i,nm} - B_{i,m}B_{j,n})
\end{aligned}
\end{equation}
Rearranging the terms we have\\
\[B_{j}^{2}\theta = B_{j}B_{j,m}(B_{i,nk} - B_{i, kn}) + B_{j}(B_{j,n}B_{i,km} - B_{j,k}B_{i,nm}) + B_{j,m}(B_{i,n}B_{j,k} - B_{i,k}B_{j,n})\]\\
Also, we have\\
\[B^{j}B_{j}\delta_{p}^{j}B_{j}(\Gamma_{jn}^{p}B_{i,mk} - \Gamma_{jk}^{p}B_{i,mn}) = B^{2}(B_{j,n}B_{i,mk} - B_{j,k}B_{i,mn})\]\\
and\\
\[B^{j}\delta_{p}^{j}B^{2}_{j}B_{j,m}(\Gamma_{in}^{j}\Gamma_{jk}^{p} - \Gamma_{ik}^{j}\Gamma_{jn}^{p}) = B^{j}B_{j,m}(B_{i,n}B_{j,k} - B_{i,k}B_{j,n})\]\\
Therefore, the revised form of equation (\ref{11}) is given as\\
\begin{equation}
\begin{aligned}
&\quad 2B^{2}B_{j}[\partial_{m}\{R_{i~kn}^{j}\}B_{j} - \partial_{m}\{R_{i~kn}^{p}\}B_{p}] + 2B^{2}B_{j}[\{R_{i~kn}^{j}\}B_{j,m} - \{R_{i~kn}^{p}\}B_{p,m}]\\
&\quad + B^{2}B_{j}(B_{j,n}B_{i,mk} - B_{j,k}B_{i,mn}) + 2B^{j}B_{j,m}(B_{i,n}B_{j,k} - B_{i,k}B_{j,n})\\
&\quad - B^{2}B_{j}(B_{j,n}B_{i,km} - B_{j,k}B_{i,nm}) - B^{j}B_{j,m}(B_{i,n}B_{j,k} - B_{i,k}B_{j,n}) = 0 \label{14}
\end{aligned}
\end{equation}
Again, as we have seen before\\
\[(B_{i,nk} - B_{i, kn}) = (2R_{i~kn}^{j} + \eta B_{j})\]\\
where, $\eta B_{j} = B_{j;kn} - B_{j;nk}$. And, we know the relation for the contravariant vector as\\
\[B_{~;ik}^{n} - B_{~;ki}^{n} = R_{i~kl}^{n}B^{l}\]\\
Now since, the metric tensor is invariant with respect to the covariant derivative, lowering the index of the vector we obtain\\
\[B_{l;ik} - B_{l;ki} = R_{i~kl}^{n}B_{n}\]\\
Thus\\
\[(B_{i,nk} - B_{i, kn}) = 2R_{i~kn}^{j}B_{j} + R_{k~ij}^{q}B_{q}\]\\
Using this relation we finally derive the {\it curvature evolution equation} as follows:
\begin{equation}
\begin{aligned}
&\quad 2B^{2}B_{j}[\partial_{m}\{R_{i~kn}^{j}\}B_{j} - \partial_{m}\{R_{i~kn}^{p}\}B_{p}] + 2B^{2}B_{j}[\{R_{i~kn}^{j}\}B_{j,m} - \{R_{i~kn}^{p}\}B_{p,m}]\\
&\quad + B^{2}B_{j}B_{j,n}(2R_{i~km}^{j}B_{j} + R_{k~ij}^{q}B_{q}) + B^{2}B_{j}B_{j,k}(2R_{i~mn}^{j}B_{j} + R_{m~ij}^{q}B_{q})\\
&\quad  + B^{j}B_{j,m}(B_{i,n}B_{j,k} - B_{i,k}B_{j,n}) = 0\\
 \label{15}
\end{aligned}
\end{equation}
The mathematical significance is immediately apparent. The physical significance will be manifest in the subsequent parts of the paper. Let us consider the special case where $k = n$. In such a scenario, the preceding equation reduces to\\
\begin{equation}
2B^{2}B_{j}[\partial_{m}\{R_{i}^{j}\}B_{j} - \partial_{m}\{R_{i}^{p}\}B_{p}] + 2B^{2}B_{j}[\{R_{i}^{j}\}B_{j,m} - \{R_{i}^{p}\}B_{p,m}]
+ B^{2}B_{j}B_{j,n}B_{q}(R_{n~ij}^{q} + R_{m~ij}^{q}) = 0 \label{16}
\end{equation}
\\
Again, $R_{i}^{p} = \delta_{j}^{p}R_{i}^{j}$ and $B_{p} = \delta_{p}^{j}B_{j}$. Also, taking into consideration another special case: $n = q = m$, we have\\
\begin{equation}
2B^{2}B_{j}^{2}\partial_{m}R_{i}^{j}[1 - \delta_{j}^{p}\delta_{p}^{j}] + 2B^{2}B_{j}B_{j,m}R_{i}^{j}[1 - \delta_{j}^{p}\delta_{p}^{j}]
+ B^{2}B_{j}B_{m}B_{j,m}R_{ij} = 0 \label{17}
\end{equation}
which is another form of the evolution equation (\ref{15}), with $m = n = q = k$. Now, if the ansatz breaks down and $j = p$  then we have from (\ref{17})\\
\[\delta_{jm}\delta_{ij}B_{j,m}B^{2}R_{ij}B^{i}B^{j} = 0\]\\
where, $R_{s} = R_{ij}B^{i}B^{j}$ is the Raychaudhuri scalar. So\\
\[B_{j,m}B^{2}R_{s} = \partial_{m}(B^{2}R_{s}) - B^{2}\partial_{m}R_{s} = 0\]\\
\[\Rightarrow B^{2} = const.\]
Assuming that this constant term doesn't change the structural form and properties of $R_{s}$ we can write without loss of any generality\\
\begin{equation}
B^{2}R_{s} \sim R_{s} \label{18}
\end{equation}
which can be looked upon as an endomorphism of the set of Raychaudhuri scalar, in this particular cosmology (with respect to the index $j$), given as\\
\begin{equation}
B^{2}:R_{s} \longmapsto R_{s} \label{19}
\end{equation}
which implies\\
\[B^{2} \circ R_{s} = R_{s}\]\\
So, when the ansatz breaks down it corresponds to the consideration of an endomorphism. On the other hand, we assumed earlier that when an endomorphism breaks down the ansatz will hold. This substantiates the plausibility of why we had associated the breakdown of an endomorphism with the ansatz we introduced. Essentially, these two can be considered to be correlated and complementary.\\
Another interpretation with regard to the endomorphism is that $B_{j}$ is a timelike {\it unit} vector field with the respect to the index $j$, which insinuates that the Raychaudhuri scalar precludes all vector fields the self scalar products of which are not constant.\\
Now, let us get back to equation (\ref{15}). Using the Kronecker delta it can be rewritten as\\
\begin{equation}
\begin{aligned}
&\quad 2B^{2}B_{j}^{2}\partial_{m}\{R_{i~kn}^{j}\}[1 - \delta_{j}^{p}\delta_{p}^{j}] + 2B^{2}B_{j}B_{j,m}\{R_{i~kn}^{j}\}[1 - \delta_{j}^{p}\delta_{p}^{j}]\\
&\quad + B^{2}B_{j}B_{j,n}(2R_{i~km}^{j}B_{j} + R_{k~ij}^{q}B_{q}) + B^{2}B_{j}B_{j,k}(2R_{i~mn}^{j}B_{j} + R_{m~ij}^{q}B_{q})\\
&\quad  + B^{j}B_{j,m}(B_{i,n}B_{j,k} - B_{i,k}B_{j,n}) = 0\\
 \label{20}
\end{aligned}
\end{equation}
Again, we know that the Brouwer's fixed point theorem \cite{Milnor} states: For any continuous function $f$ mapping a compact convex set to itself there is a fixed point.\\
Therefore, considering a continuous mapping $f$ in our Riemannian manifold and a compact, geodesically convex vector field ($\mathfrak{B}$) comprised of timelike vectors $B_{i}$, we shall have\\
\[f: \mathfrak{B} \longmapsto \mathfrak{B}\]\\
and a fixed point under this automorphism. Incidentally, choosing the index $j$ we can infer that there is a fixed point with respect to this index in the vector field (or the geodesic field), through which the family of vectors $B_{j}$ are parallel transported. Consequently, the vectors $B_{j}$ will be constant irrespective of the coordinates and the geometric structure of the manifold. Thus, equation (\ref{20}) becomes\\
\[2B^{2}B_{j}^{2}\partial_{m}\{R_{i~kn}^{j}\}[1 - \delta_{j}^{p}\delta_{p}^{j}] = 0\]
\begin{equation}
R_{i~kn}^{j} = const. \label{21}
\end{equation}
Hence, the curvature tensor returns to its initial state with respect to the index $j$. This corresponds to the statement of Poincare's recurrence theorem \cite{Poincare}, if one considers the whole manifold to be a system.\\
Now, let us consider equation (\ref{17}). Writing, $1 - \delta_{j}^{p}\delta_{p}^{j} = \delta$, we have\\
\[2B^{2}B^{2}_{j}\delta\delta^{jk}\partial_{m}R_{ik} + 2B^{2}B_{j}B_{j,m}\delta\delta^{jk}R_{ik}
+ B^{2}B_{j}B_{m}B_{j,m}\delta_{j}^{k}R_{ik} = 0\]\\
or,\\
\[2B^{2}B^{2}_{j}\delta^{jk}\partial_{m}R_{ik} + 2B^{j}B_{j,m}\delta_{ji}R_{ik}B^{i}B^{k}
+ B^{2}B_{j,m}\delta_{j}^{k}\delta_{ji}\delta_{mk}R_{ik}B^{i}B^{k} = 0\]\\
Since, we have considered the case where $j \neq p$, $\delta = 1$. So, using the expression for the Raychaudhuri scalar ($R_{s}$) and after rearranging we have finally\\
\begin{equation}
B_{j,m}R_{s} = -\rho B^{2}B_{j}^{2}R_{ik,m} \label{22}
\end{equation}
where, $\rho = 2\delta^{jk}(2\delta_{ji}B_{j}^{-1} + \delta_{j}^{k}\delta_{ji}\delta_{mk})^{-1} = 2\delta^{jk}(2\delta_{ji}B_{j}^{-1} + \delta_{mi})^{-1}$. Thus, we have obtained the result that under certain conditions the Raychaudhuri scalar depends on the derivative of the Ricci tensor ($R_{ik}$), the associated timelike vector field ($B_{j}$) and its derivatives.\\
Since, it is known that $R_{s}$ is the trace of the tidal tensor epitomizing the relative acceleration due to gravity of two objects separated by an infinitesimal distance and that $R_{ik}$ measures the change in geometry as an object moves along geodesics in the space, we can conclude: For some particular timelike vector field in a Riemannian or pseudo-Riemannian manifold, the relative acceleration due to gravity decreases with the increase in curvature and vice versa.

\section{Sectional curvature}
In this section, we analyze and discuss the results of the preceding section. Firstly, let us consider the parameter $\rho$ of equation (\ref{22}). Now, for $j = k \neq i$ and $m = i$ we have\\
\begin{equation}
B_{j,i}R_{s} = -2B^{2}B_{j}^{2}R_{ij,i} \label{23}
\end{equation}
where, $R_{s} = R_{ij}B^{i}B^{j}$. On the other hand, for $i = j = k$, $i \neq m$ we have with $R_{s} = R_{jj}B^{j}B^{j}$. Hence, we shall obtain\\
\begin{equation}
B_{j,m}R_{s} = -B^{2}B_{j}^{3}R_{jj,m} \label{24}
\end{equation}
Now, equation (\ref{23}) can also be written as\\
\begin{equation}
B_{j,i}R_{s} = -B^{2}B_{j}^{2}[R_{ij,i} + R_{ij,i}] \label{25}
\end{equation}
Now, if we take into consideration the automorphism introduced in the previous section, then\\
\[B_{j,i} = 0\]\\
Consequently, we have\\
\begin{equation}
R_{ij,i} + R_{ij,i} = 0 \label{26}
\end{equation}
The general implication of this equation is that the Ricci curvature, $R_{ij}$, is constant with respect to the index $i$. However, there can be another implication that the first term gives a positive value and the second gives a negative value - a notion that might seem erroneous, but can be used as an alternative explanation. To be precise\\
\begin{equation}
|R_{ij,i}| = \pm  R_{ij,i} \label{27}
\end{equation}
which essentially insinuates that\\
\begin{equation}
|R_{ij}| = \pm R_{ij} \label{28}
\end{equation}
i.e. the Ricci curvature can have both positive and negative values. Again, since $R_{ij}$ is obtained by contracting the RC tensor which in turn is related to the sectional curvature of a Riemannian manifold (with respect to the given manifold and two linearly independent tangent vectors at the same point), we can conclude that the sectional curvature also will have both positive and negative values. This bespeaks for both geodesic convergence and divergence on account of Rauch's comparison theorem \cite{Rauch1,Rauch2} that states: for positive sectional curvature, geodesics tend to converge and for negative sectional curvature, geodesics tend to diverge.\\
Essentially, under special circumstances, geodesics can diverge. Geodesic convergence leads to a singularity; similarly geodesic divergence would lead to a point of exclusivity as shown previously by resorting to Raychaudhuri's equation. \\
It is interesting to point out that during the inflationary era, curvature played a significant role pertinent to the dynamics, as researchers have found out \cite{Ali}.

\section{Discussions}
In the present article, we have established that an equation epitomizing the evolution of curvature, resorting to an ansatz originating from the breakdown of an endomorphism correlated to the Raychaudhuri scalar. It is also shown that the ansatz and the endomorphism are interrelated in the sense that one precludes the other.\\
From the aforementioned considerations we have also found that the Riemann-Christoffel curvature tensor tends to follow Poincare's recurrence theorem and thereby the curvature returns to its initial value after certain period of time.\\
Another interesting consequence is that the existence and feasibility of negative curvature which entails geodesic divergence. This is also validated by using the expansion scalar which is show to have both positive and negative values. This negative value and that of the Riemann-Christoffel tensor indicates that there might exist a point of \emph{exclusivity} which is the opposite of the point of \emph{singularity} - a result that has been derived from the Raychaudhuri equation too.\\
Ostensibly, the notion of such a point is in a sense a speculative extrapolation and demands ample amount of study and research. But, the prospect is something worth investigating, at the very least.\\


\begin{thebibliography}{99}
\bibitem{Einstein} A. Einstein, \emph{The Foundation of the General Theory of Relativity}, Annalen der Physik. 354 (7): 769, 1916. 
\bibitem{Karl} K. Schwarzschild, \emph{Über das Gravitationsfeld eines Massenpunktes nach der Einsteinschen Theorie}, Sitzungsberichte der Königlich Preussischen Akademie der Wissenschaften. 7: 189–196, 1916.
\bibitem{Friedmann1} A. Friedmann, \emph{Uber die Krümmung des Raumes}, Zeitschrift fur Physik A, 10 (1): 377–386, 1922.
\bibitem{Friedmann2} A. Friedmann, \emph{Uber die Moglichkeit einer Welt mit konstanter negativer Krümmung des Raumes}, Zeitschrift fur Physik A, 21 (1): 326–332, 1924.
\bibitem{Lemaitre1} G. Lemaitre, \emph{Expansion of the universe, A homogeneous universe of constant mass and increasing radius accounting for the radial velocity of extra-galactic nebulae}, Monthly Notices of the Royal Astronomical Society, 91 (5): 483–490, 1931.
\bibitem{Lemaitre2} G. Lemaitre, \emph{l'Univers en expansion}, Annales de la Societe Scientifique de Bruxelles, A53: 51–85, 1933.
\bibitem{Robertson1} H. P. Robertson, \emph{Kinematics and world structure}, Astrophysical Journal, 82: 284–301, 1935.
\bibitem{Robertson2} H. P. Robertson, \emph{Kinematics and world structure II}, Astrophysical Journal, 83: 187–201, 1936.
\bibitem{Robertson3} H. P. Robertson, \emph{Kinematics and world structure III}, Astrophysical Journal, 83: 257–271, 1936.
\bibitem{Walker} A. G. Walker, \emph{On Milne's theory of world-structure}, Proceedings of the London Mathematical Society, Series 2, 42 (1): 90–127, 1937.
\bibitem{Kerr} R. P. Kerr, \emph{Gravitational field of a spinning mass as an example of algebraically special metrics}, Physical Review Letters. 11 (5): 237–238, 1963.
\bibitem{Ezra} E. Newman et. al., \emph{Metric of a Rotating, Charged Mass}, Journal of Mathematical Physics. 6 (6): 918–919, 1965.
\bibitem{Raychaudhuri} A. K. Raychaudhuri, \emph{Relativistic cosmology I}, Phys. Rev. 98 (4): 1123–1126, 1955.
\bibitem{Penrose1} R. Penrose, \emph{Gravitational collapse and space-time singularities}, Phys. Rev. Lett., 14 (3): 57, 1965.
\bibitem{Penrose2} S. Hawking \& R. Penrose, \emph{The Singularities of Gravitational Collapse and Cosmology}, Proceedings of the Royal Society A. 314 (1519): 529–548, 1970. 
\bibitem{Das} A. Das, \emph{A Hamiltonian formalism for the universe and its implications}, IJMPA, Vol. 33, No. 29, 1850171 (2018).
\bibitem{Caldwell} R. R. Caldwell et. al., \emph{Phantom Energy and Cosmic Doomsday}, Physical Review Letters. 91 (7): 071301, 2003.
\bibitem{Jayant} Jayant V. Narlikar, \emph{An Introduction to Cosmology}, Cambridge University Press, pages 50-57, ISBN 0521793769, 2002.
\bibitem{Milnor} J. Milnor, \emph{Analytic proofs of the "hairy ball theorem" and the Brouwer fixed-point theorem}, Amer. Math. Monthly 85, no. 7, pp. 521-524, 1978.
\bibitem{Poincare}  H. Poincare, \emph{Sur le probleme des trois corps et les equations de la dynamique}, Acta Math. 13: 1–270, 1890.
\bibitem{Rauch1} H. E. Rauch, \emph{A contribution to differential geometry in the large}, Ann. Math. 54 (1): 38–55, 1951.
\bibitem{Rauch2} M. P. do Carmo, \emph{Riemannian Geometry}, Birkhauser, 1992.
\bibitem{Ali} Ali A. Asgari \& Amir H. Abbassi, \emph{Imprints of Intrinsic and Exterior Curvatures in Cosmology}, Gravitation and Cosmology, volume 27, pages 152–156, 2021.
\end{thebibliography}
\end{document}